# SOFTWARE AND FIRMWARE-LOGIC DESIGN FOR THE PIP-II MACHINE PROTECTION SYSTEM MODE AND CONFIGURATION CONTROL AT FERMILAB*

L. Carmichael, M. Austin, J. Eisch, E. Harms, R. Neswold, A. Prosser, A. Warner, J. Wu
Fermilab, Batavia, IL 60510, U.S. A

*Abstract*
The PIP-II Machine Protection System (MPS) requires a dedicated set of tools for configuration control and management of the machine modes and beam modes of the accelerator. The protection system reacts to signals from various elements of the machine according to rules established in a setup database in the form of a Look-Up-Table filtered by the program Mode Controller. This is achieved in accordance with commands from the operator and governed by the firmware logic of the MPS. This paper describes the architecture, firmware logic, and implementation of the program mode controller.

## INTRODUCTION

The Proton Improvement Plan-II (PIP-II) is an enhancement to the Fermilab accelerator complex [1] that will provide intense high energy neutrino beam to the Deep Underground Neutrino Experiment (DUNE) [2]. PIP-II (Fig.1) will consist of a 800 MeV H-Superconducting linac which includes a Warm Front-end (WFE), and a 300-meter-long beam transfer line to the Fermilab Booster. The WFE of the linac plays a critical role in the accelerator. It generates a 30 KeV H- beam, defines the beam parameters, accelerates the beam to an energy of 2.1 MeV with its RFQ for compatibility with downstream accelerating structures, and generates a required bunch pattern. One of the high-level goals of the machine is to deliver a proton beam power to target in excess of 1 MW with sustained high reliability along with multiuser operations of the Fermilab complex. A PIP-II Injector Test Facility [3] (PIP2IT) was assembled in multiple stages in 2014-2021 as a testbed for PIP-II technologies and protection schemes.

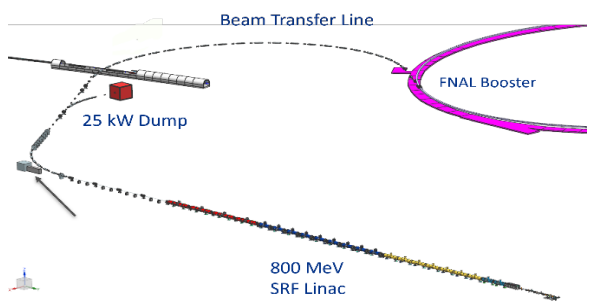

Figure 1: Schematic of PIP2 Facility

A Machine Protection System (MPS) capable of managing high power is needed to protect the accelerator and its components from beam induced damage. This system also needs to allow for seamless transition from the defined operational modes while maintaining MPS integrity. Four Beam Modes specify the beam pulse length and repetition rate, and seven Machine Modes specify beam path configurations. In each Beam Mode the machine can be in one of three states – Standby, Ready or Beam on. The required fast response time to shut off the linac beam in response to critical detected faults and beam losses above measured thresholds is 10 microseconds, accomplished by inhibiting the Low Energy Beam Transport (LEBT) line Chopper which is one of four (4) Beam Inhibit Devices (BIDs). Commissioning of the WFE is scheduled for 2026 with phased commissioning of the full linac thereafter. The final design of the MPS have been approved and is commensurate with this timescale.

## MPS SYSTEM ARCHITECTURE

The system architecture is shown in Figure 2. The MPS is FPGA based and consists of a Main MPS (MPSM) which issues the system permits and interfaces with the BIDs, an Analog MPS (MPSA) for post-processing of digitized signals derived from certain beam current measuring devices and a Digital MPS (MPSD) which processes serialized inputs from machine subsystems coming from the field via serializer hardware.

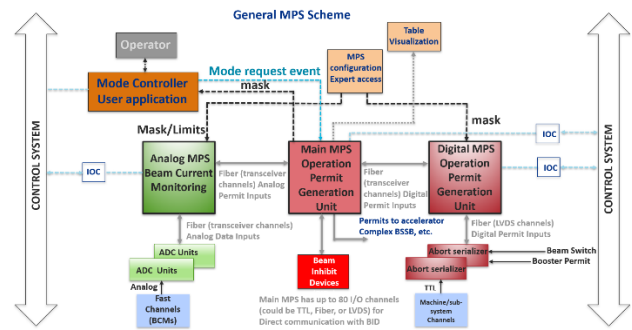

Figure 2: MPS Integrated Overview Diagram

**The MPSM** receives all permit input status bits from the MPSA and MPSD Units. For expansion purposes it can accommodate up to 20 fiber transceiver channels from multiple MPSA and MPSD units. It can send data to other MPS units, and it can accept up to 80 I/O channels for direct connection with the BIDs. These can be TTL, Fiber or LVDS. Based on the Operation Mode or Masked channel configuration the system will read all the inputs from the MPSD and MPSA units and decide whether to disable or establish certain BID permits. Depending on how, or why a BID permit was removed there is also the ability to manually or automatically clear or reset the permits based Operating Mode, the clock events received, or control system interaction over the controls network.

The **MPSA** units are designed to handle up to 40 analog channels from a total of 10 ADC units that are connected to 20 fiber transceiver channels in the field. The units have all respective settings, (min, max, etc.) for the various operating Modes associated with each channel so that when the Operating Mode event and data is decoded the system can use the respective settings for comparison with digitized analog values received from the ADC units. With this information the MPSA determines whether to disable the permit input for a given channel to the MPSM or keep it alive. The analog MPS also sends this signal back to its respective ADC unit as a status output. For differential measurements where analog channels will be compared with each other prior to a permit decision, the channels are deliberately connected to the same MPSA Unit. If more than 20 channels are needed to be compared, then MPSAs can be linked to each other to transfer data between them. This adds a finite amount of time to the permit decision process.

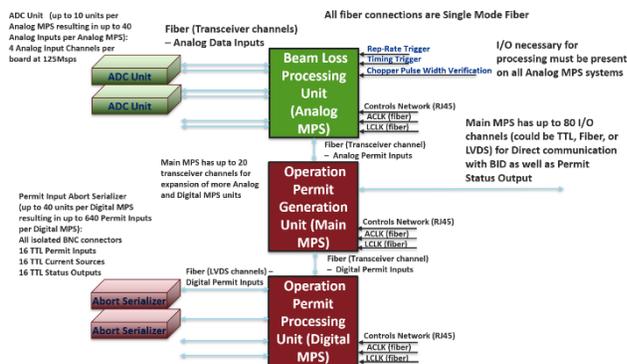

Figure 3: Interfaces among MPS Units

The **MPSD** units are designed to accommodate up to 40 fiber Tx/Rx LVDS channels and connect to serializer boxes in the field. Serializers take 16 independent TTL level inputs from subsystems (This results in 640 inputs per MPSD Unit) and internally encode them into 20 bits using an 8-bit/10-bit (8B/10B) encoding format. The resulting 20-bit word is then transmitted differentially at 20 times the reference clock rate or 600Mbps to the MPSD Unit. This connection is via a Fiber LVDS connection. These serializers provide expandability and consolidation of input signals throughout various segments of the PIP-II accelerator. The serializer are designed to be as simple as possible and therefore do not contain any programmable chips for universal ease of swap outs regardless of location. All the respective settings (HI, LOW, masked, etc.) in the MPSD for each channel in the various Operating modes will reside in a Look-Up-Table for making fast decisions. When the Operating Mode EVENT and data is decoded, the respective Setting is compared with the Input Permits received from the Serializers. The MPSD Unit determines whether to disable the Permit Input for that channel to the MPSM or keep it active. It also sends this signal back to its respective Serializer as a Status Output. Similar to the MPSAs, in the case where Digital channels will be compared with other Digital channels prior to deciding to disable a permit, then those channels will be placed in the same MPSD.

## MPS FIRMWARE LOGIC AND DESIGN

The operating mode within the MPSM and MPSA will be specified and set by an 'Operating Mode Event' – a timing event generated by the accelerator clock/timing system. This allows for a synchronous changing of the modes at a known time and to broadcast the current Operating Mode to the accelerator complex. The mode Event will be played out at a known time before the 'beginning of beam Pulse' Events. This provides the window for changing operating modes at rates compatible with hardware and data rate specification. The permit inputs to the MPS are updated about every 33 nanoseconds. This mechanism also provides timing that accommodates High-Level-Applications which are designed to interact with machine operationally to accomplish beam measurements and/or studies that may involve invasive devices or tuning.

There is a Look-Up-Table (LUT) within the MPS that maps each independent Operating Mode to a set of inputs a priori to determine the beam permit. The mode currently selected determines which inputs are needed for beam.
Since the MPSA has all possible settings (min, max, etc.) for each analog channel the currently selected mode determines which values are used for calculating whether a channel is kept valid or not. This valid/invalid value for each analog channel is sent from the MPSA to the MPSM where it is then treated in the same fashion as the permit inputs into the MPSM and compared with the LUT. Permit inputs as well as analog channel setting that have known constant values for specific Operating Modes are hardcoded for that mode into the MPS and are not settable from the network to avoid operator error. Settings for a specific Operating Mode's values (masked permit inputs, analog ranges, etc.) are only be accepted as settings into the MPS (both Main and Analog) when the 'Operating Mode' Event is played, and the data value is set to a "No-Beam" mode. At that point, all user settable settings can be entered into the MPS. After settings have been made, the desired Operating Mode value is placed into the 'Operating Mode' EVENT where it is played out on the clock and the MPS receives this clock EVENT which begins operation in that selected mode with the current settings all at once.

## BEAM INTERRUPT AND RECOVERY

There are four beam inhibit devices (BIDs) in the accelerator located in the Ion Source (IS) and the LEBT Figure 4. There are the LEBT chopper, the Modulator

Extractor, the IS High Voltage Supply (ISHV), and the LEBT Dipole. These are divided into two tiers, BID-1 and BID-2. These two tiers are used to prioritize and classify certain types of faults and provide some automatic recovery for system faults such as rf interruptions. When such faults occur, the beam is first interrupted by only BID-1 devices and a fault counter for the channel is incremented by 1. As soon as the offending channel starts to report OK status, the beam is allowed by the next coming pulse. However, if the fault is still reported after 10 μs, the failure to stop the beam with only the BID-1 devices is reported and BID-2 devices are instructed to disable the beam permit. If the fault counter reaches the maximum number specified in the MPS (e.g.,10), all BIDs are instructed to disable the beam. If no new fault occurs in the time specified in the beam permit tables (e.g., 1 min), the counter is reset to zero.

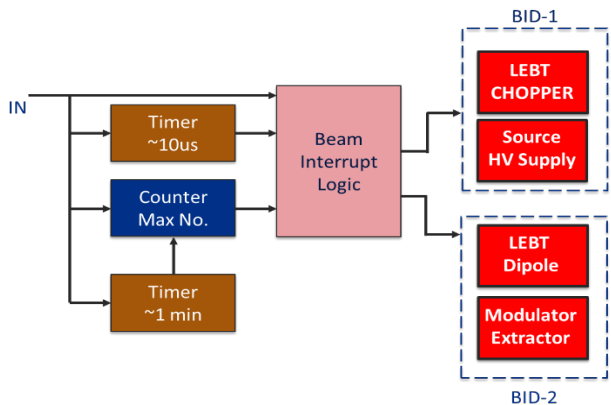

Figure 4: Primary and Secondary Device Diagram

## USER INTERFACE

The user interface under development for the MPS will allow operators and domain experts to interact with the MPS via a set of applications. These applications include Phoebus displays for viewing faults, java application for configuration control and flutter applications for post-mortem analysis.

The underlying controls framework is EPICS with IOCs handling the readback and control of the MPS FPGA boards. Additionally, a middle tier is being used which consists of a set of services called Data Pool Managers (DPM). These services interact not only with the EPICS IOCs, but also with legacy front ends. Also, these services provide interfaces to Java, Python and Web frameworks being used by several of the user interface applications. This layer is illustrated in Fig. 5.

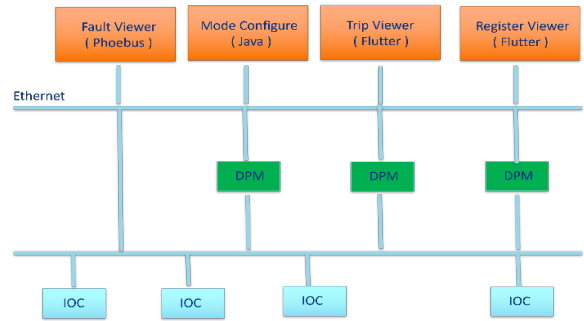

Figure 5: DPM Middle Tier

The overall effect is to provide operators and experts a variety of tools to display and analyse faults and provide configuration control for the MPS.

## HIERARCHICAL FAULT VIEWER

One of the main MPS applications is the Fault Viewer Phoebus display. This display provides users with operator and expert views of faults, as shown in Fig. 6

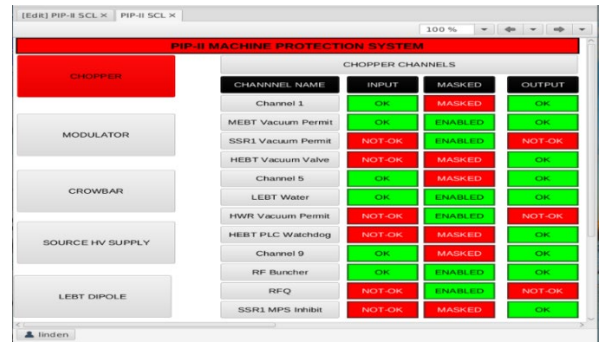

Figure 6: Hierarchical Fault Viewer

This view provides some coarse granularity in identifying faulted channels and tripped BIDs. The user can then drill down to a specific BID to view the channels that are producing a fault.

## POST-MORTEM ANALYSIS

Post-mortem analysis is a key component of the MPS workflow. The user interface is an essential component of the MPS. Once a trip is detected and the permit removed, it is necessary to be able to drill down to find the source of this trip and then to be able to reconfigure your system to handle future trips.

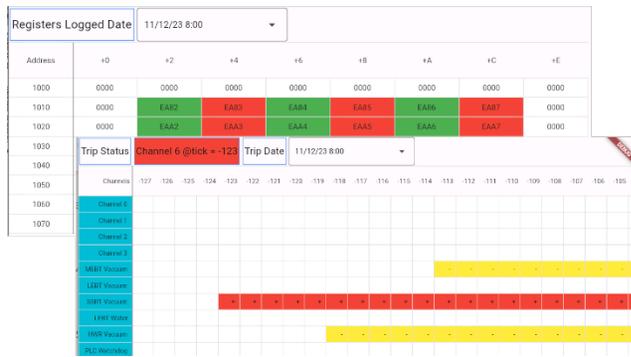

Figure 7: Trip and Register Displays

The above figure illustrates 2 flutter applications, the trip monitor and the register monitor. These applications interface IOCs which timestamp, and log detected trips and their associated channels. on the MPS FPGA.

The trip monitor application allows users to retrieve trip information associated with a particular trip. This information includes which channels faulted and the order and timestamp associated with each channel fault. This information is also logged, and users can retrieve past trips to find fault trends.

The register monitor application is a similar flutter application. It provides the user with a snapshot of the registers on the FPGA. This is useful to the domain experts as it allows them to analyse the state of the MPS at specific instances. This information is also logged and retrievable.

## MODE CONFIGURATION AND CONTROL

A distinct group of channels is monitored to generate a permit for each BID. The decision to monitor a channel can change depending on the beam type and path of the beam. A set of Beam modes and Machine modes were defined that capture this dependency and are illustrated in Figure 8.

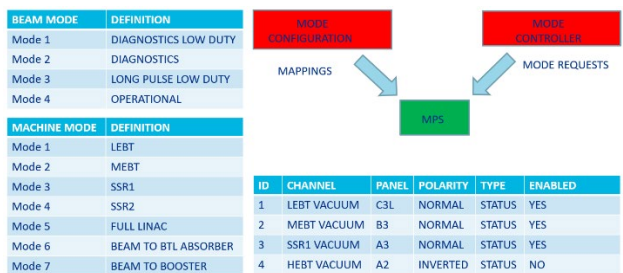

Figure 8: Machine and Beam Modes

A mapping was developed which maps Beam and Machine modes to a distinct set of channels for each BID. The configuration control process starts with domain experts saving a distinct set of enabled channels for each Beam and Machine mode combination on the MPS FPGA. This mapping is done separately for each BID. This mapping is only updated if new channels are added or removed. Operators issue a request for a new Beam and Machine mode. This request is sent to the MPS FPGA which then enables the channels mapped to requested Beam and Machine modes.

The Mode Configuration Application is a java application interfaced to the Data Pool Manager middleware developed at Fermilab. This application allows experts to save mappings between modes and MPS parameters, masks, and limits on the MPS.

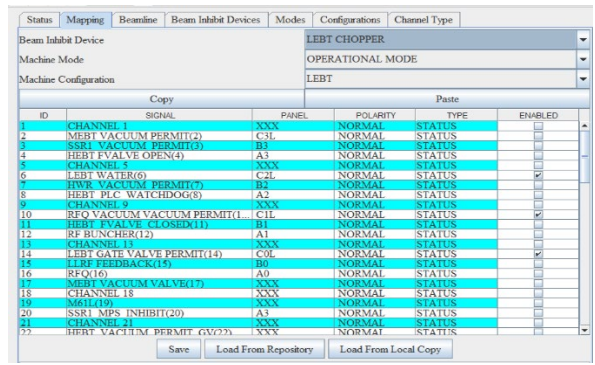

Figure 9: Mode configuration application

The mode Controller Application is a Phoebus display which allows operators to send a mode change request to the MPS which will then switch to the masks and limits mapped to the requested Beam and Machine modes.

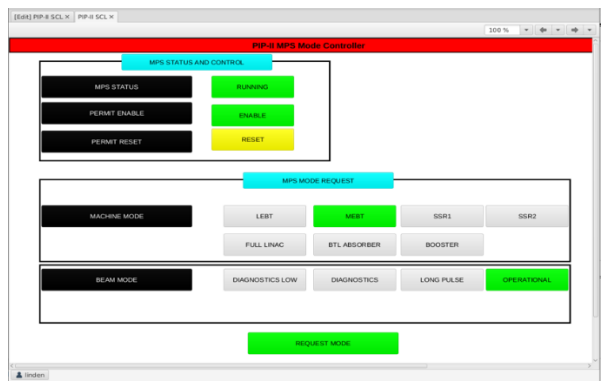

Figure 10: Mode control Application

Authentication is a key component of the MPS workflow and is essential to reducing human error. Users are broken into different groups and assigned different privileges. Authentication is being explored via the Keycloak framework which uses identity and access management to handle authentication. The intent is to expand this to include role-based authentication.

# CONCLUSION

The PIP-II Machine Protection System has undergone final review approvals and is in the build and development phase. The focus of the design efforts was concentrated on an architecture that centralizes the permit decision processes into a Main MPS. The system includes the operational logic at the firmware level where possible and mode and some system configuration changes will be affected through the timing and clock system. At a higher software level, mode and configuration management and user interfaces are being designed that are based on EPICS. The design considers future expansion of the system to accommodate upgrades and operation knowledge. A framework for MPS operations with High-Level-Applications is also envisioned and the timing system will be used to broadcast the machine state to such application.